\documentclass[10pt, journal]{IEEEtran}
\usepackage{cite}
\usepackage{bm}
\usepackage{amsmath}
\usepackage{extarrows}
\usepackage{amssymb}
\usepackage{graphicx}
\usepackage{color}
\usepackage{enumerate}
\usepackage{bookmark} 
\usepackage{booktabs}
\graphicspath{{figure}}
\usepackage{setspace}
\usepackage{subfigure}
\usepackage{algorithm}
\usepackage{algorithmicx}
\usepackage{multirow}
\usepackage{stfloats}
\usepackage{graphbox}
\usepackage{bbm}

\newcommand{\mr}{\mathrm}

\newcommand{\BE}{\begin{equation}}
\newcommand{\EE}{\end{equation}}
\newcommand{\BS}{\begin{subequations}}
\newcommand{\ES}{\end{subequations}}
\renewcommand{\bf}{\bm}

\newcommand{\tabincell}[2]{\begin{tabular}{@{}#1@{}}#2\end{tabular}}

\allowdisplaybreaks \allowdisplaybreaks[2]

\ifCLASSINFOpdf
\graphicspath{{figure/}}

\else

\fi

\begin{document}
\bstctlcite{BSTcontrol}

\title{{Interleave Frequency Division Multiplexing}}

\author{\IEEEauthorblockN{Yuhao Chi, \emph{Senior Member, IEEE}, Lei Liu*, \emph{Senior Member, IEEE}, Yao Ge, \emph{Member, IEEE}, \\ Xuehui Chen, Ying Li, \emph{Member, IEEE}, and Zhaoyang Zhang, \emph{Senior Member, IEEE}}

\thanks{Yuhao Chi, Xuehui Chen, and Ying Li are with the State Key Laboratory of Integrated Services Networks, School of Telecommunications Engineering, Xidian University, Xi'an, 710071, China (e-mail: yhchi@xidian.edu.cn, xuehui\_chen@stu.xidian.edu.cn, yli@mail.xidian.edu.cn).}

\thanks{Lei Liu and Zhaoyang Zhang are with the Zhejiang Provincial Key Laboratory of Information Processing, Communication and Networking, College of Information Science and Electronic Engineering, Zhejiang University, Hangzhou, 310007, China. (e-mail: \{lei\_liu, ning\_ming\}@zju.edu.cn). (\emph{*Corresponding author: Lei Liu)}}
\thanks{Yao Ge is with the Continental-NTU Corporate Lab, Nanyang Technological University, Singapore 639798 (e-mail: yao.ge@ntu.edu.sg).}
}

\maketitle
\begin{abstract}
In this letter, we study interleave frequency division multiplexing (IFDM) for multicarrier modulation in static multipath and mobile time-varying channels, which outperforms orthogonal frequency division multiplexing (OFDM), orthogonal time frequency space (OTFS), and affine frequency division multiplexing (AFDM) by considering practical advanced detectors. {The fundamental principle underlying existing modulation techniques is to establish sparse equivalent channel matrices in order to facilitate the design of low-complexity detection algorithms for signal recovery, making a trade-off between  performance and implementation complexity. In contrast, the proposed IFDM establishes an equivalent fully dense and right-unitarily invariant channel matrix with the goal of achieving channel capacity, ensuring that the signals undergo sufficient statistical channel fading. Meanwhile, a low-complexity and replica maximum \emph{a posteriori} (MAP)-optimal cross-domain memory approximate message passing (CD-MAMP) detector is proposed for IFDM by exploiting the sparsity of the time-domain channel and the unitary invariance in interleave-frequency-domain channel.} Numerical results show that IFDM with extremely low-complexity CD-MAMP outperforms OFDM, OTFS, and AFDM with state-of-the-art orthogonal approximate message passing detectors, particularly at low velocities.
\end{abstract}

\vspace{-0.3cm}
\section{Introduction}
With the rapid development of high-mobility communication scenarios, orthogonal frequency division multiplexing (OFDM) in 5G struggles to ensure reliable data transmission due to severe inter-carrier interference caused by large Doppler shifts. To address this issue, orthogonal time frequency space (OTFS) is developed~\cite{OTFS1}, multiplexing information symbols in the delay-Doppler (DD) domain to mitigate channel delays and Doppler shifts. Recently, affine frequency division multiplexing (AFDM) is proposed~\cite{Afdmtwc}, which uses the inverse discrete affine Fourier transform (IDAFT) to modulate information symbols into a ``warped" time-frequency domain, demonstrating full diversity in high-mobility channels. {Although the diversity gain can be exploited by OTFS and AFDM with extremely complex maximum likelihood (ML) detectors, this does not ensure capacity-optimal performance. The essence of OFDM, OTFS, and AFDM is the construction of sparse equivalent channel matrices to enable low-complexity signal detection algorithms for a trade-off between optimal performance and implementation complexity.}




Many low-complexity detectors for OTFS and AFDM are currently being investigated. For OTFS, a cross-domain orthogonal approximate message passing (CD-OAMP) detector is developed in \cite{OTFS-OAMP}, which is iteratively performed between the time domain and the DD domain. A DD-OAMP has been proposed in the DD domain for OTFS \cite{OTFS_DDOAMP}. However, because CD/DD-OAMP employs linear minimum mean-square error (LMMSE), the channels' sparsity is not fully exploited due to the channel inverse, resulting in high complexity. To address this issue, a DD-domain memory approximate message passing (DD-MAMP) detector is presented~\cite{MAMPOTFSconf}, employing a memory matching filter (MF) to utilize the DD channels' sparsity.  Nonetheless, DD-MAMP ignores the sparser time-domain channels. The development of AFDM detectors is still in its early stages.  For example, {simplified linear detectors are investigated in \cite{Afdmtwc,AfdmLow}, limiting the detection performance. There is still a scarcity of efficient AFDM detectors.}

{This letter investigates an interleaved frequency division multiplexing (IFDM) technique with the goal of achieving channel capacity, in which all symbols undergo sufficient statistical  fading based on an equivalent fully dense channel matrix obtained by a randomly interleaved inverse Fourier transform, termed as the IF transform. For brevity, the capitals ``I'' and ``F'' in IF denote the interleave and Fourier, respectively. Meanwhile, an extremely low-complexity replica MAP-optimal CD-MAMP detector is proposed by fully exploiting super-sparse time-domain channels.} Numerical results show that IFDM with CD-MAMP outperforms OFDM, OTFS, and AFDM with state-of-the-art OAMP in MIMO at $300$ km/h by more than $3$ dB, especially at low velocity. For MIMO-IFDM with 512 subcarriers, CD-MAMP approaches OAMP's BER performance with a 100-fold time reduction.

{
The major contributions of this paper are summarized:
\begin{enumerate}
    \item The investigated IFDM significantly outperforms OFDM, OTFS, and AFDM by enhancing the statistical stability of static multipath and mobile time-varying channels when considering practical advanced detectors.
    \item The proposed CD-MAMP achieves replica MAP-optimal\footnote{{To date, the replica MAP/capacity optimality has been recognized as an optimality metric, where MMSE and constrained capacity predicted by the replica method are proved to be correct for independently and identically distributed Gaussian matrices\cite{ReevesTIT2019} , and certain specific sub-classes of unitarily invariant matrices\cite{zhoufan}. A rigorous proof of the replica method for a wider range of unitarily invariant matrices remains an open issue.}} performance with extremely low complexity by using the IF transform to enable the equivalent channel matrix satisfying the right-unitarily invariant assumption.
\end{enumerate}
}

{{\emph{Related Works:}} Similar concepts are found in \cite{OFDM_post_inte, EST_Li,EST_yuan,xisheng2023vector}. In \cite{OFDM_post_inte}, an interleaver is added after the IFFT of OFDM to introduce time diversity against impulse noise. The energy-spread transform (EST) was first developed in \cite{EST_Li} for channel equalization. A turbo-based equalization algorithm is proposed for EST-based MIMO in~\cite{EST_yuan}. In \cite{xisheng2023vector}, EST-based probabilistic constellation shaping is designed for multipath channels with soft frequency domain equalization using the vector approximate message passing algorithm. Surprisingly, significant advantages have already been demonstrated in \cite{EST_yuan,xisheng2023vector}. Coincidentally, our findings reveal for the first time that IFDM surpasses existing OFDM, OTFS, and AFDM in static multipath and mobile time-varying channels with practical advanced detectors. Meanwhile, the proposed CD-MAMP detector can maximally exploit channel sparsity as well as achieve replica MAP-optimal performance in IFDM.}

\begin{figure*}[t!]\vspace{-0.5cm}
	\centering
	\includegraphics[width=0.75\textwidth]{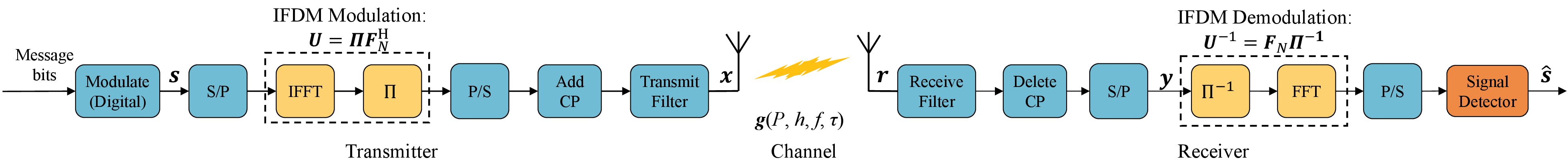}\vspace{-0.2cm}
	\caption{An IFDM-based multicarrier communication system, where $\bf{\Pi}$ and $\bf{\Pi}^{-1}$ denotes random interleaving and deinterleaving, $\bf{F}^{\mr{H}}$ and $\bf{F}$ denote an IFFT matrix and an FFT matrix, respectively, and S/P and P/S denote serial-to-parallel and parallel-to-serial conversion, respectively.}\label{Fig:sysm}\vspace{-0.3cm}
\end{figure*}

\vspace{-0.2cm}
\section{Preliminaries}
\subsection{OFDM Modulation}
A symbol vector $\bf{s}\in \mathbb{C}^{N\times 1}$ in the frequency domain is modulated by the IFFT to generate a time-domain signal $\bf{x}$, 
\BE\label{Eqn:ofdm}
\bf{x} = \bf{F}_{N}^{\mr{H}}\bf{s},
\EE
where $\bf{F}_{{N}}$ denotes the $N$-point normalized FFT matrix. Then, a  cyclic prefix (CP) is added to $\bf{x}$ and transmitted over static multipath or mobile time-varying channels.

After OFDM demodulation, the received signal is
\BE\label{Eqn:rev_ofdm}
\bf{y} = \bf{F}_{N}(\bf{H}\bf{x}+\bf{w}) = \bf{F}_{N}(\bf{F}_{N}^{\mr{H}}\bf{\Lambda}\bf{F}_{N}\bf{x}+\bf{w})=\bf{\Lambda}\bf{s}+\tilde{\bf{w}},
\EE
where $\bf{H}$ is a circulant channel matrix for static multipath channels and its eigenvalue decomposition is $\bf{H}=\bf{F}_{N}^{\mr{H}}\bf{\Lambda}\bf{F}_{N}$ with frequency-domain channel response $\bf{\Lambda}$,  $\bf{w}\sim \mathcal{CN}(\bf{0},\sigma^2\bm{I})$ is additive white Gaussian noise (AWGN), and $\tilde{\bf{w}}\triangleq \bf{F}_{N}\bf{w}$. However, due to Dopplers in time-varying channels, $\bf{H}$ is no longer a circulant matrix, thus, its eigenvalue decomposition in \eqref{Eqn:rev_ofdm} does not hold, resulting in severe inter-carrier interference.

\vspace{-0.4cm}
\subsection{OTFS Modulation}
To address the performance loss of OFDM in mobile time-varying channels, information symbols $\bf{S}\in \mathbb{C}^{K\times L}$ on a two-dimensional (2D) $K \times L$ grid in the DD domain of OTFS are transformed into a time-domain signal vector $\bf{x}\in \mathbb{C}^{N\times 1}$ by the ISFFT and Heisenberg transform (HT). That is, 
\BE\label{Eqn:ofts}
\bf{x}
=\mr{vec}(\overbrace{\bf{I}_{K}\bf{F}_{K}^{\mr{H}}}^{\text{
HT}}\overbrace{\bf{F}_{K}\bf{S}\bf{F}_{L}^{\mr{H}}}^{\text{ ISFFT}})
=(\bf{F}_{L}^{\mr{H}}\otimes \bf{I}_{K})\cdot{\mr{vec}}(\bf{S}),
\EE
where $N=KL$, {$\bf{I}_{K}$ is the $K$-dimensional identity matrix,} $\bf{F}_{K}\in \mathbb{C}^{K\times K}$ and $\bf{F}_{L}\in \mathbb{C}^{L\times L}$ are the $K$-point and $L$-point normalized FFT matrices. After adding the CP, $\bf{x}$ is transmitted over the time-varying channels $\bf{H}$.

The received signal after OTFS demodulation is 
\BE\label{Eqn:revofs}
\begin{aligned}
\bf{y} &= (\bf{F}_{L}\otimes \bf{I}_{K})(\bf{H}\bf{x}+\bf{w})\\
&= (\bf{F}_{L}\otimes \bf{I}_{K})\bf{H}(\bf{F}_{L}^{\mr{H}}\otimes \bf{I}_{K}){\mr{vec}}(\bf{S}) +
\tilde{\bf{w}},
\end{aligned}
\EE
where {$\bf{H}_{\mr{eff}} \triangleq (\bf{F}_{L}\otimes \bf{I}_{K})\bf{H}(\bf{F}_{L}^{\mr{H}}\otimes \bf{I}_{K})$ is sparse by discarding small-valued elements, especially for fractional Doppler cases, } and $\tilde{\bf{w}}\triangleq(\bf{F}_{L}\otimes \bf{I}_{K})\bf{w}$.

\subsection{AFDM Modulation}
Unlike OTFS, a symbol vector $\bf{s}\in \mathbb{C}^{N\times 1}$ in AFDM is multiplexed by the IDAFT into time-domain signals, i.e., 
\BE\label{Eqn:afdm}
\bf{x}
=\overbrace{\bf{\Lambda}_{c_1}^{\mr{H}}\bf{F}^{\mr{H}}_{{N}}\ \bf{\Lambda}_{c_2}^{\mr{H}}}^{\text{IDAFT}}\bf{s}=\bf{A}^{-1}\bf{s},
\EE
where $\bf{A}\triangleq \bf{\Lambda}_{c_2}\bf{F}_{{N}}\bf{\Lambda}_{c_1}$ with $\bf{\Lambda}_{c_i}\triangleq\mr{diag}(e^{-j2\pi c_in^2}, n=0, ..., N-1)$, $i=1,2$. Then, a chirp-periodic prefix (CPP)~\cite{Afdmtwc} is added to $\bf{x}$ and transmitted over time-varying channels $\bf{H}$.

Following AFDM demodulation, the received signal is 
\BE\label{Eqn:rvafdm}
\bf{y}=\bf{A}(\bf{H}\bf{x}+\bf{w}) = \bf{A}\bf{H}\bf{A}^{-1}\bf{s} + \tilde{\bf{w}},
\EE
where {$\bf{H}_{\mr{eff}}\triangleq \bf{A}\bf{H}\bf{A}^{-1}$ is sparsified by ignoring many elements with smaller values \cite{Afdmtwc}} and $\tilde{\bf{w}}\triangleq\bf{A}\bf{w}$.

\vspace{-0.5cm}
\section{Interleave Frequency Division Multiplexing}

\subsection{IFDM Modulation and Demodulation}
Fig.~\ref{Fig:sysm} shows an IFDM-based single-input-single-output (SISO) system. A message bit sequence is digitally modulated as $\bf{s} \in \mathbb{C}^{N \times 1}$ with the power constraint $\frac{1}{N}{||\bf{s}||^2}=1$, the elements of which are individually taken from constellation set $\mathcal{S}$ (e.g., QPSK, QAMs). Following serial-to-parallel (S/P) conversion and IFDM modulation, the obtained signal $\bf{x}$ is 
\BE\label{Eqn:smatix}
\bf{x}=\bf{U}\bf{s} = \bf{\Pi}\bf{F}^{\mr{H}}_N\bf{s},
\EE
where $\bf{U}\triangleq \bf{\Pi}\bf{F}^{\mr{H}}_N\in \mathbb{C}^{N\times N}$ denotes the IF transform with the random interleave matrix $\bf{\Pi} \in \mathbb{R}^{N\times N}$. Noting that $\bf{U}$ is a unitary matrix,  $\bf{U}\bf{U}^{\mr{H}}=\bf{I}$. To cope with inter-frame interference and multipath propagation, a CP of length at least equal to the maximum channel delay spread is added to $\bf{x}$. After the transmit filter, the signal $\bf{x}$ is sent out.

The received signal ${r}[n]$ at the $n$-th slot is given by
\BE\label{Eqn:Rev}
r[n]=\textstyle\sum\nolimits_{p=0}^{P-1}x[n-p]g[n,p]+w[n], 
\EE
with the equivalent baseband channel impulse response 
\BE\label{Eqn:gn}
g[n, p] = \textstyle\sum\nolimits_{i=1}^{L}h_ie^{j2\pi f_i(nT_s -pT_s)}\mr{P_{rc}}(pT_s-\tau_i),
\EE
where $g[n, p]$ is assumed to be available only to the receiver, $p=0, ..., P-1$, $P$ is the channel tap, $L$ is the number of multipaths, $T_s$ is the system sampling interval, and $\{h_i, f_i, \tau_i\}$ represent the complex gain, the Doppler shift, and the delay at the $i$-th path, respectively. Let $\tau_{\text{max}} \triangleq \text{max}(\tau_i) < N$ and $P$ is determined by $\tau_{\text{max}}$ as well as the duration of the overall filter response. $\mr{P_{rc}}(\cdot)$ is the overall raised-cosine rolloff filter when {the practical root raised-cosine (RRC) pulse shaping filters are employed at the transceiver to control signal bandwidth and reject out-of-band emissions.} 

After the receiver filter and CP removal, \eqref{Eqn:Rev} is written as
\BE\label{Eqn:rev_mat}
\bf{y} ={\bf{H}}\bf{x}+\bf{w},
\EE
where ${\bf{H}}\in \mathbb{C}^{N\times N}$ denotes the effective time-domain channel. Followed by inverse IF, the received signal is given by
\BE\label{Eqn:rev_v}
\hat{\bf{y}}=\bf{U}^{-1}\bf{y}
=\bf{F}_N\bf{\Pi}^{-1}\bf{H}\bf{\Pi}\bf{F}_N^{\mr{H}}\bf{s}+\bf{w}
=\bf{H}_{\mr{eff}}\bf{s}+\tilde{\bf{w}},
\EE
where $\bf{H}_{\mr{eff}}\triangleq \bf{F}_N\bf{\Pi}^{-1}\bf{H}\bf{\Pi}\bf{F}_N^{\mr{H}}$, $\tilde{\bf{w}}\triangleq \bf{U}^{-1}\bf{w}$, and $\bf{\Pi}^{-1}$ is the inverse random interleave matrix. Based on \eqref{Eqn:rev_v}, common signal detection algorithms can be employed. 


\begin{figure}[t]\vspace{-0.4cm}
	\centering
	\includegraphics[width=0.8\columnwidth]{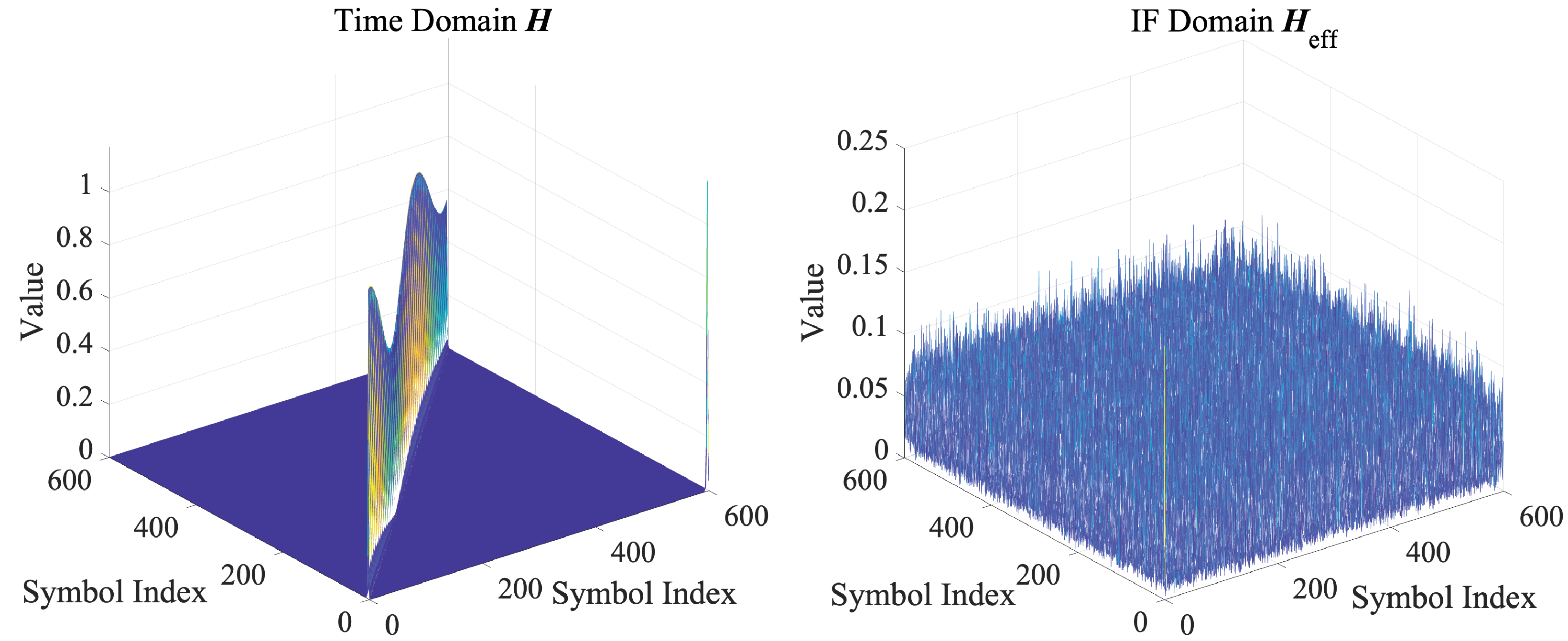}\vspace{-0.1cm}
	\caption{Comparison of time domain $\bf{H}$ and IF domain ${\bf{H}}_{\mr{eff}}$.}\label{Fig:Channel}\vspace{-0.2cm}
\end{figure}


{\emph{Notes:} An intuitive interpretation of the key principle of IFDM is the utilization of the IF transform to  make $\bf{H}_{\mr{eff}}$ statistically more stable, as shown in Fig. \ref{Fig:Channel}, which is conducive to signal recovery. The reason is that all signals are assigned to all subcarriers at random and independently, enabling all signals undergo statistical channel fading.}
Meanwhile, {the IF transform enables the equivalent channel matrices to satisfy the right-unitarily invariant assumption, which is commonly used for OAMP and MAMP algorithms, ensuring general practical OAMP and MAMP detectors achieve replica MAP-optimal performance\cite{LeiMAMP} and channel capacity with proper code design \cite{YufeiTWC2024}. {Note that replica MAP/capacity optimality is potentially diversity-optimal, while diversity optimality is not guaranteed to be MAP/capacity optimal.} 
Table~\ref{Tab:Modu} summarizes the comparison of OFDM, OTFS, AFDM, and IFDM. Note that IFDM has a modulation complexity of $\mathcal{O}(N{\mr{log}}N)$, which is the same as that of OFDM and AFDM~\cite{Afdmtwc} but less than that of OTFS~\cite{OTFS-OAMP}.  Given the CD/DD-OAMP detector\cite{OTFS-OAMP,OTFS_DDOAMP} or the low-complexity CD-MAMP detector proposed later, IFDM is the best in terms of BER, while OFDM is the worst.


\vspace{-0.4cm}
\subsection{MIMO-IFDM System}
Fig. \ref{Fig:mimosysm} shows that IFDM is integrated with MIMO. Modulated signal $\bf{s}$ through S/P and IFDM modulation is divided into $N_t$ segments, and CP is added to each segment. After the transmit filter, signal $\bf{x}_j$ is obtained and transmitted through the channel at the $j$-th antenna, $j=1, ..., N_t$.
Here, the channel impulse response between the $j$-th transmit antenna and the $m$-th receive antenna is expressed as
\BE\nonumber 
g_{m,j}[n,p]\!\!= \!\!\textstyle\sum\nolimits_{i=1}^{L_{m,j}}\!\!h_{m,j,i}e^{j2\pi f_{m,j,i}(nT_s -pT_s)}\mr{P_{rc}}(pT_s-\tau_{m,j,i}),
\EE
where $n=0, ..., N-1$, $m=1, ..., N_r$, $p=0, ..., P_{m,j}-1$, $L_{m,j}$ is the number of multipaths between the $j$-th transmit antenna and the $m$-th receive antenna, and $\{h_{m,j,i}, f_{m,j,i}, \tau_{m,j,i}\}$ denote the channel gain, the Doppler shift, and the delay associated with the $i$-th path, respectively. 

\begin{table}[t] \scriptsize \vspace{-0.5cm}
 \caption{Comparisons of OFDM, OTFS, AFDM, and IFDM.}\label{Tab:Modu}
 \vspace{-0.2cm}
 \centering\setlength{\tabcolsep}{1mm}{
\begin{tabular}{|c|c|c|c|c|}
\hline
Modulation & \tabincell{c}{Transform\vspace{-0.15cm}\\Domain} &\tabincell{c}{Modulation\vspace{-0.15cm}\\Complexity} & \tabincell{c}{Equivalent Channel \vspace{-0.1cm}\\Matrix Type} & BER  Rank\\
\hline
\multirow{2}{*}{OFDM} & \multirow{2}{*}{Frequency} & \multirow{2}{*}{$\mathcal{O}(N\mr{log}N)$} & Diagonal (static) & \multirow{2}{*}{3 (worst)}\\ 
\cline{4-4}
 & & &  Dense (mobile) & \\ 
\hline
OTFS & Delay-Doppler & $\mathcal{O}(\frac{3}{2} N\mr{log}N)$ & Sparse & \multirow{2}{*}{2}\\ 
\cline{1-4}
AFDM& DAFT & {$\mathcal{O}(N\mr{log}N)$} & Sparse  &  \\  
\hline
\tabincell{c}{{IFDM}\vspace{-0.15cm}\\({proposed})} & IF & $\mathcal{O}(N\mr{log}N)$ & Random Dense & {\textbf{1 (best)}} \\
\hline
\end{tabular}
}
\end{table}
\begin{figure}[t]\vspace{-0.3cm}
	\centering
	\includegraphics[width=0.7\columnwidth]{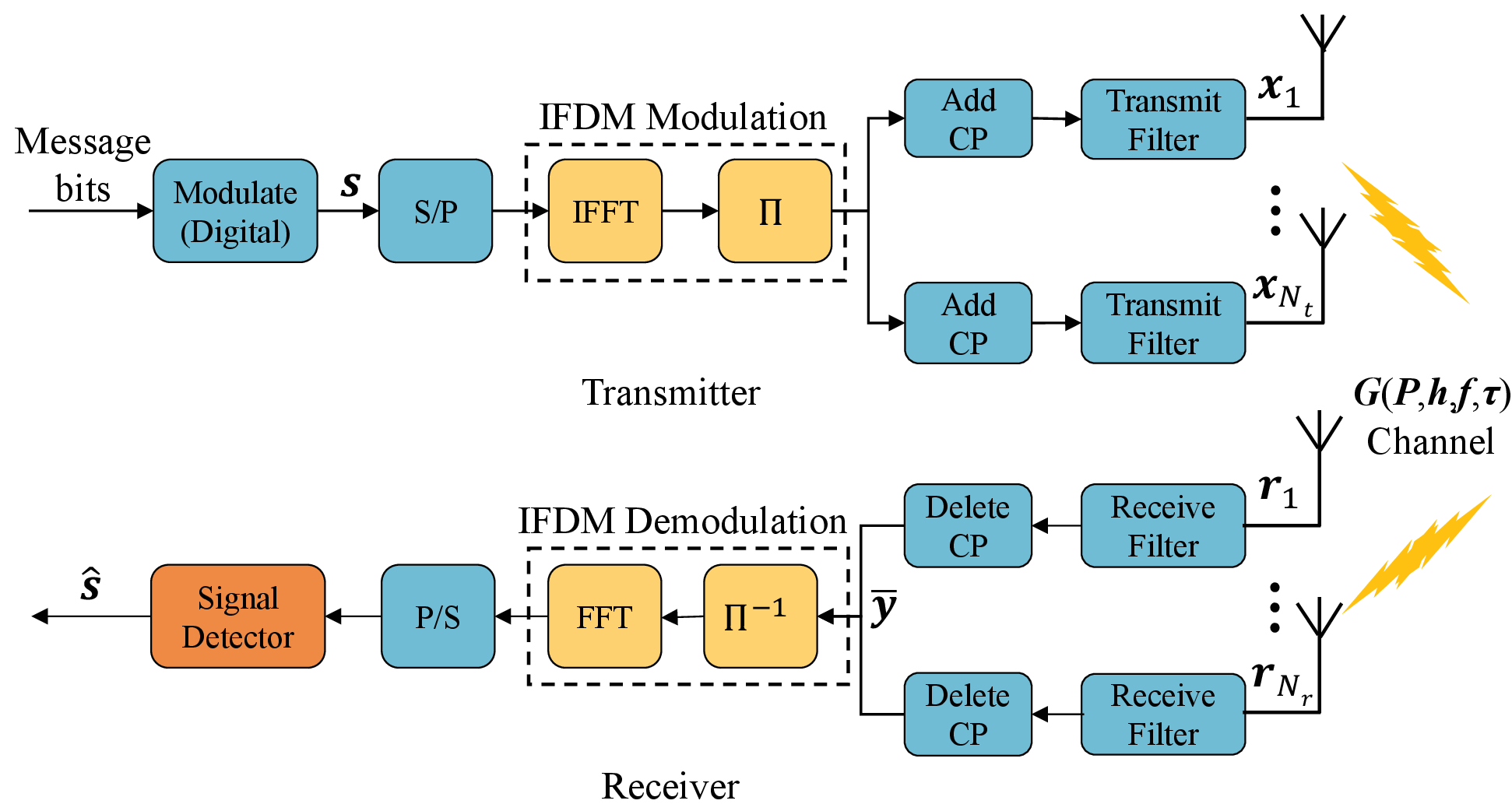}\vspace{-0.3cm}
	\caption{MIMO-IFDM  with $N_t$-antenna transmitter and $N_r$-antenna receiver.}\label{Fig:mimosysm}\vspace{-0.2cm}
\end{figure}

The received signal $\bf{r}_m$ at the $m$-th receive antenna first passes through the receive filter. After discarding the CP, the signals is expressed as $\bar{\bf{y}}=[\bar{\bf{y}}_1^{\mr{T}}, ..., \bar{\bf{y}}_{N_r}^{\mr{T}}]^{\mr{T}}\in \mathbb{C}^{N_rN\times 1}$, i.e.,
\BE \label{Eqn:ifdm_mimo}
\bar{y}_m[n]=\textstyle\sum\limits_{j=1}^{N_t}\textstyle\sum\limits_{p=0}^{P_{m,j}-1} g_{m,j}[n,p]x_{j}[[c-p]_{N}]+\bar{w}_{m}[n].
\EE
Thus, $\bar{\bf{y}}$ is rewritten as
\BE\label{Eqn:ifdm_mimo2}
\bar{\bf{y}} = \bar{\bf{H}}\bar{\bf{x}}+\bar{\bf{w}},
\EE
where $\bar{\bf{x}}\!=\![\bf{x}_1^{\mr{T}}, ..., \bf{x}_{N_t}^{\mr{T}}]^{\mr{T}}\in \mathbb{C}^{N_tN\times 1}$, {$\bar{\bf{H}}\!=[\bf{H}_1^{\mr{T}}, ...,$ $ \bf{H}_i^{\mr{T}}, ..., \bf{H}_{N_r}^{\mr{T}}]^{\mr{T}}\in \mathbb{C}^{N_rN\times N_tN}$, $\bf{H}_i\in\mathbb{C}^{N\times N_tN}$, $i= 1, ..., N_r$}, and $\bar{\bf{w}}\!\!=\!\![\bar{\bf{w}}_{1}^{\mr{T}}, ..., \bar{\bf{w}}_{N_r}^{\mr{T}}]^{\mr{T}}\in \mathbb{C}^{N_rN\times1}$.
Following IFDM demodulation, signal $\bf{s}$ is recovered using signal detection algorithms.

It is worth noting that the existing modulations seek to sparsify the equivalent channels\cite{OTFS-OAMP,OTFS_DDOAMP,MAMPOTFSconf,Afdmtwc}, facilitating the use of low-complexity detectors at the receiver for signal recovery. {This is a trade-off between performance and implementation complexity.} In contrast, the IF transform in IFDM ensures that each transmission symbol undergoes enough statistical channels, but it results in full-dense equivalent channel matrices. As a result, the most difficult challenge of IFDM is to develop a low-complexity and replica MAP-optimal detector. In the next section, we will propose a low-complexity and replica MAP-optimal CD-MAMP to address this problem perfectly.

\vspace{-0.2cm}
\section{Cross-domain MAMP detector}
In this section, we propose a low-complexity CD-MAMP detector composed of a memory MF in the time domain and a nonlinear detection (NLD) in the IF domain, as shown in Fig.~\ref{Fig:mamp_rev}. The memory MF achieves extremely low complexity by multiplying the sparse time-domain channel matrix with the input signals. The complete CD-MAMP is given below and can be directly extended to MIMO-IFDM.

\subsubsection{Memory~MF}
Based on \eqref{Eqn:rev_mat} and \emph{a-priori} information from NLD, i.e., starting with $t=1$ and $\bf{X}_1=\bf{0}$,
\BE\label{Eqn:LMMF}
\bf{r}_{t} = \gamma_t(\bf{X}_t) = \tfrac{1}{\varepsilon^{\gamma}_t}(
\hat{\gamma}_{t}(\bf{X}_t)-\bf{X}_t\bf{p}_t ), 
\EE
where $\bf{X}_t=[\bf{x}_1,...,\bf{x}_t]$ and $\hat{\gamma}_{t}(\bf{X}_t)=\bf{H}^{\rm{H}}\tilde{\gamma}_t(\bf{X}_t)$ with
\BE\label{Eqn:MLd}
\tilde{\gamma}_t(\bf{X}_t)=\bf{B}_t \tilde{\gamma}_{t-1}(\bf{X}_{t-1}) + \xi_t(\bf{y}-\bf{H}\bf{x}_t),
\EE
$\tilde{\gamma}_0(\bf{X}_0)=\bf{0}$ and $\bf{B}_t=\theta_t (\lambda^{\dagger}\bf{I}-\bf{H}\bf{H}^{\rm{H}})$. {Noting that $\gamma_t(\bf{X}_t)$ is a memory MF that depends mainly on $\bf{H}$ and all previously estimations $\bf{X}_{t}=[\bf{x}_1,..., \bf{x}_t]$, where $\bf{x}_t$ is the output of the damping operation between $[\bf{x}_1,..., \bf{x}_{t-1}]$ and $\tilde{\bf{x}}_{t}$, i.e.,
\BE\label{Eqn:damp1}
{\bf{x}}_{t} = [{\bf{x}}_1, ..., {\bf{x}}_{t-1}, \tilde{\bf{x}}_{t}]\cdot \bf{\zeta}_{t},
\EE
with {damping vector $\bf{\zeta}_{t}=[\zeta_{t, 1}, ..., \zeta_{t, t}]^{\text{T}}$} and  $\tilde{\bf{x}}_{t}$ from the NLD.}  {$\bf{\zeta}_{t}$ is used for the linear weighted superposition of all estimations, which is used to guarantee and improve the convergence of MAMP in principle. The parameters $\{\varepsilon^{\gamma}_t, \bf{p}_t, \xi_t, \theta_t, \lambda^{\dagger}\}$ are adjusted to ensure convergence and replica MAP optimality of MAMP with right-unitarily invariant matrices (See details in \cite{LeiMAMP}).}

\begin{figure}[t!]\vspace{-0.5cm}
	\centering
	\includegraphics[width=0.9\columnwidth]{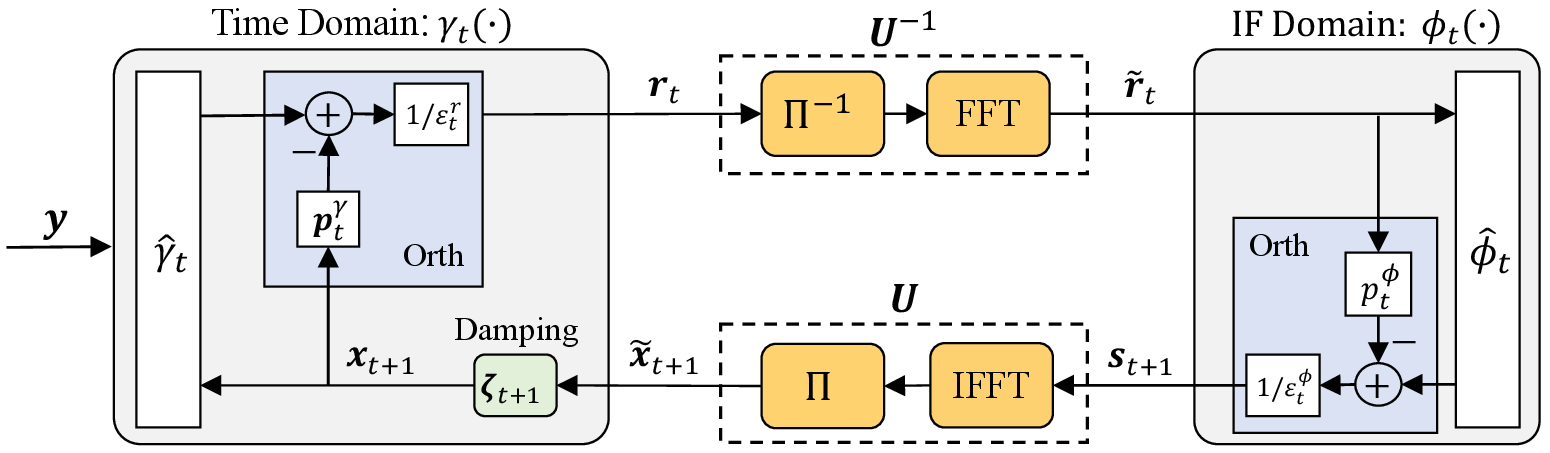}\vspace{-0.2cm}
	\caption{The CD-MAMP detector in IFDM: memory MF detector $\gamma_t(\cdot)$ in the time domain and MMSE nonlinear detector $\phi_t(\cdot)$ in the IF domain.}\label{Fig:mamp_rev} \vspace{-0.4cm}
\end{figure}
Based on \eqref{Eqn:LMMF}, the output estimate variance of $\gamma_t(\cdot)$ is 
\BE\label{Eqn:le_var1}
v_{t,t}^{\gamma} \equiv \tfrac{1}{N}\mr{E}\{||\gamma_t(\bf{X}_t)-\bf{x}||^2\}.
\EE

Note that the estimation error of the current output is orthogonal to the estimation errors of all preceding inputs \cite{LeiMAMP}. This enables the asymptotically independent identically distributed (IID) Gaussianity of estimate errors, i.e., $\bf{r}_t = \bf{x} + \bf{z}_t^{\gamma}$, where $\bf{z}_t^{\gamma}$ $ \sim \mathcal{CN}(\bf{0}, v_{t,t}^{\gamma}\bf{I})$ is an AWGN noise and independent of $\bf{x}$.

\subsubsection{Inverse IF Transform} Through the inverse IF transformation $\bf{U}^{-1}$, signal $\tilde{\bf{r}}_t$ and variance $\tilde{{v}}_{t,t}^{\gamma}$ are obtained as 
\BS
\begin{align}
\tilde{\bf{r}}_t &= \bf{U}^{-1}\bf{r}_t=\bf{F}\bf{\Pi}^{-1}\bf{r}_t,  \label{Eqn:uest} \\
\tilde{{v}}_{t,t}^{\gamma}&= v_{t,t}^{\gamma}\bf{U}^{\mr{H}}\bf{U}=v_{t,t}^{\gamma}.
\end{align}
\ES

It is worth noting that the unitary property of $\bf{U}$ can enhance the IID Gaussianity of the estimation errors in $\tilde{\bf{r}}_t$, i.e., $\tilde{\bf{r}}_t=\bf{s}+\tilde{\bf{z}}_t^{\gamma}$ with $\tilde{\bf{z}}_t^{\gamma}=\bf{U}^{-1}\bf{z}_t^{\gamma}\sim \mathcal{CN}(\bf{0}, v_{t,t}^{\gamma}\bf{I})$.  Fig.~\ref{Fig:qqplot} shows that $\tilde{\bf{r}}_t$ is closer to the ideal Gaussian signal with variance $v_{t,t}^{\gamma}$, where $\bf{s}$ is an equiprobable BPSK signals, i.e., $\mr{E}\{\bf{s}\}=\mr{E}\{\bf{x}\}=\bf{0}$.


\subsubsection{NLD} The NLD $\phi_t(\cdot)$ consists of a symbol-by-symbol MMSE demodulation $\hat{\phi}_t(\cdot)$ and orthogonalization. The output estimation of $\phi_t(\cdot)$ is 
\BE\label{Eqn:nle_orth}
{\bf{s}}_{t+1} = \phi_t(\tilde{\bf{r}}_t)  = \tfrac{1}{\varepsilon_t^{{\phi}}}(\hat{\phi}_t(\tilde{\bf{r}}_t)-p_{t}^{{\phi}}\tilde{\bf{r}}_t),
\EE
where $\hat{\phi}_t(\tilde{\bf{r}}_t) = \mr{E}\{\bf{s}|\tilde{\bf{r}}_t, \bf{s} \in \mathcal{S}\}$ and $\{\varepsilon_t^{{\phi}},p_{t}^{{\phi}}\}$ are the normalization and orthogonalization parameters in \cite{LeiMAMP}. 


The corresponding output variance of $\phi_t(\cdot)$ is 
\BE\label{Eqn:nldvar2}
{v}_{t+1,t+1}^{{\phi}}\equiv \tfrac{1}{N}\mr{E}\{||{\phi}_t(\tilde{\bf{r}}_t)-\bf{s}||^2\}.
\EE

\begin{figure}[t!]\vspace{-0.5cm}
	\centering
	\includegraphics[width=0.75\columnwidth]{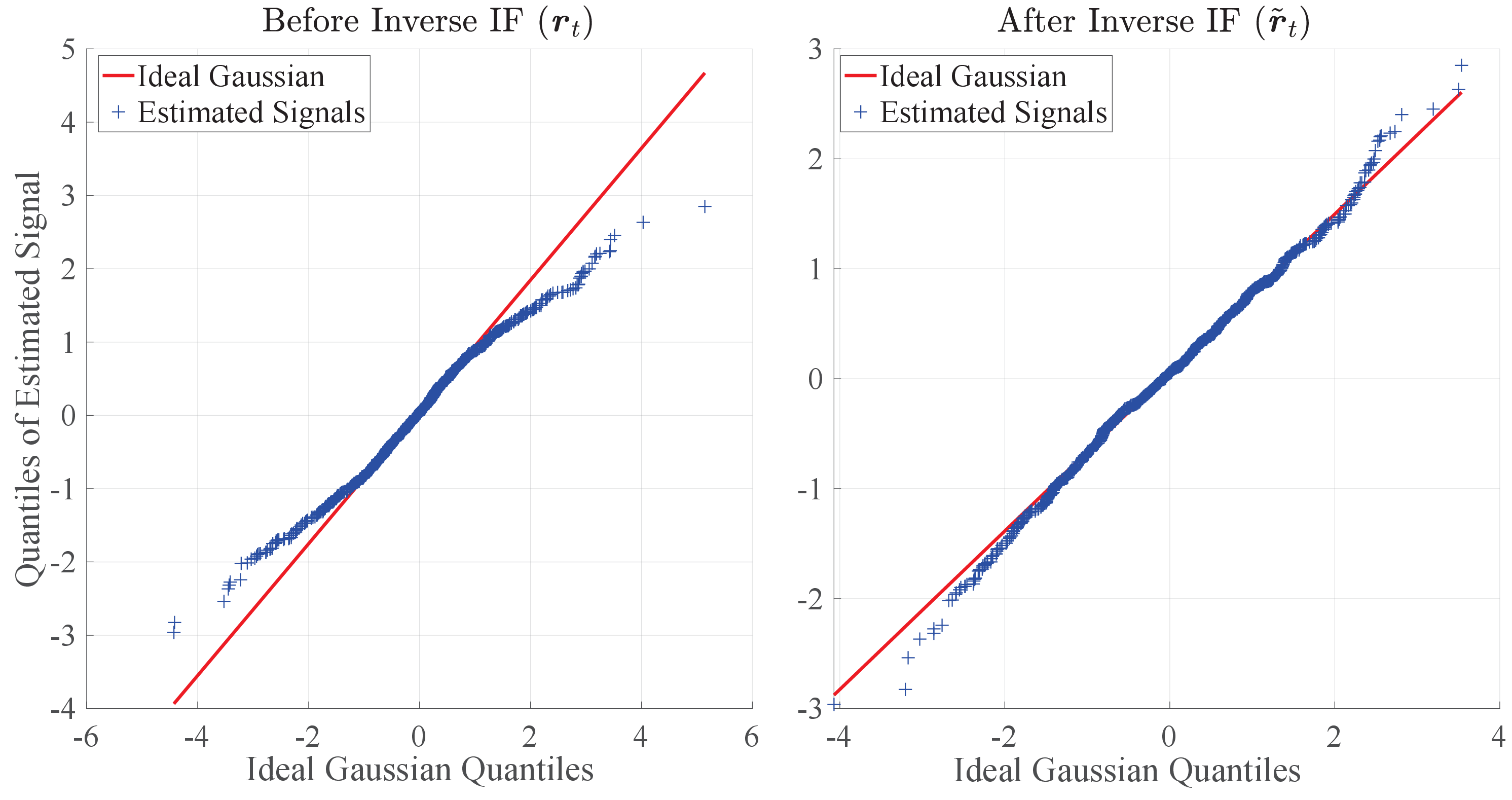}\vspace{-0.2cm}
	\caption{Quantile-quantile plot between the estimation signals before and after the inverse IF transform (i.e., $\bf{r}_t$ and $\tilde{\bf{r}}_t$) and ideal Gaussian quantiles with $\mathcal{CN}(\bf{0}, v_{t,t}^{\gamma}\bf{I})$ and $t=1$, where $\bf{s}$ is an equiprobable BPSK signal vector.}\label{Fig:qqplot} 
\end{figure}

\subsubsection{IF Transform} 
Through the IF transform $\bf{U}$, signal $\tilde{\bf{x}}_{t+1}$ and variance $\tilde{v}_{t+1,t+1}^{\phi}$ are obtained as 
\BS
\begin{align}
\tilde{\bf{x}}_{t+1} &= \bf{U}\bf{s}_{t+1}=\bf{\Pi}\bf{F}^{\mr{H}}\bf{s}_{t+1},  \label{Eqn:ifuest} \\
\tilde{v}_{t+1,t+1}^{\phi}&= {v}_{t+1,t+1}^{{\phi}}\bf{U}\bf{U}^{\mr{H}}={v}_{t+1,t+1}^{{\phi}},
\end{align}
\ES
which are fed back to the memory MF for the next iteration.

{To validate the advantages of the proposed CD-MAMP, Table~\ref{Tab:comp} presents a comparison of complexity and performance ranking with CD/DD-OAMP\cite{OTFS-OAMP,OTFS_DDOAMP}, DD-MAMP\cite{MAMPOTFSconf}\footnote{Based on \eqref{Eqn:rev_mat}, DD-OAMP/MAMP proposed in \cite{OTFS_DDOAMP,MAMPOTFSconf} can be used directly for signal detection in IFDM.}, and Gaussian message passing (GMP)\cite{OTFS_GMP} detectors, where $\mathcal{T}$ denotes the iteration number and $P \ll N$. It is obvious that CD-MAMP can approach the  BER performance of state-of-the-art CD/DD-OAMP\cite{OTFS-OAMP,OTFS_DDOAMP} with the lowest complexity.}

\begin{table}[t] \scriptsize \vspace{-0.4cm}
 \caption{Comparisons of Advanced Detectors.}\label{Tab:comp}
 \vspace{-0.2cm}
 \centering\setlength{\tabcolsep}{1mm}{
\begin{tabular}{|c|c|c|c|}
\hline
\multicolumn{2}{|c|}{Detector} & \multicolumn{2}{c|} {Performance Rank}\\ 
\hline
Algorithm & Complexity & Complexity & BER \\ \hline
LMMSE & $\mathcal{O}(N^3)  $ & 4 & {3 (worst)}\\  \hline
\tabincell{c}{GMP \cite{OTFS_GMP} (unstable,\vspace{-0.15cm}\\ 
heuristic damping)} & $\mathcal{O}(N^2\mathcal{T})$ & 3 &   2\\
\hline
{CD-OAMP \cite{OTFS-OAMP}}& $\mathcal{O}(N^3\mathcal{T}+2N\mathcal{T}{\mr{log}}N)$ & 6 (worst) & \multirow{4}{*}{\tabincell{c}{\textbf{1 (best)}\vspace{-0.1cm}\\almost same}}\\
\cline{1-3}
{DD-OAMP\cite{OTFS_DDOAMP}}& $\mathcal{O}(N^3\mathcal{T})$ & 5 &   \\ 
\cline{1-3}
{DD-MAMP\cite{MAMPOTFSconf}} & $\mathcal{O}(N^2\mathcal{T})$& 2 & \\ 
\cline{1-3}
\tabincell{c}{{CD-MAMP}\vspace{-0.15cm}\\({proposed})} &  $\mathcal{O}(PN\mathcal{T}+2N\mathcal{T}{\mr{log}}N)$ & \textbf{1 (best)} & \\  
\hline
\end{tabular}\vspace{-0.4cm}
}
\end{table}

\begin{figure*}\vspace{-0.7cm}
\centering
\subfigure[\fontsize{4.7pt}{\baselineskip}\selectfont{OFDM/OTFS/AFDM/IFDM: \!QPSK+SISO\! 
 (300km/h)}]{\label{Fig:BER_P2P_diffmod}
\includegraphics[width=0.25\linewidth]{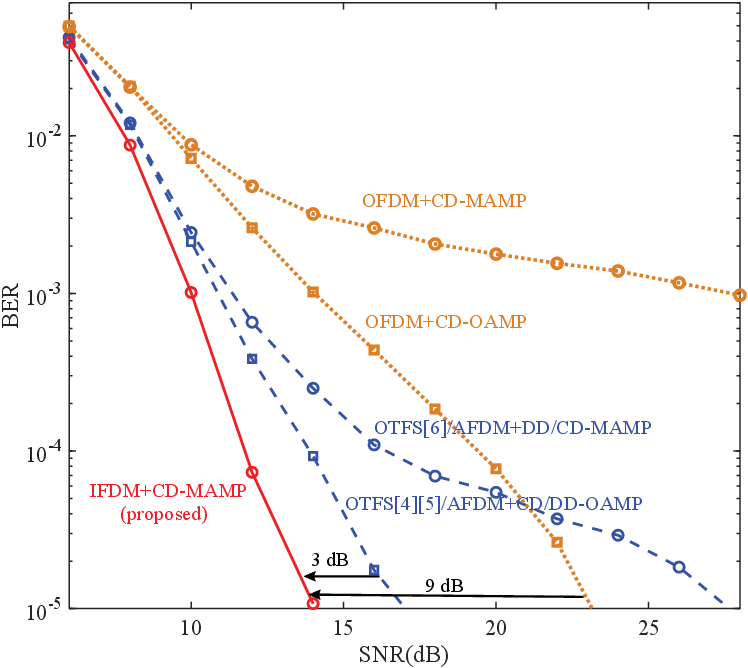}}
\subfigure[\fontsize{4.7pt}{\baselineskip}\selectfont{OFDM/OTFS/AFDM/IFDM: \!QPSK+MIMO\! 
 (300km/h)}]{\label{Fig:BER_MIMO_diffmod}
\includegraphics[width=0.245\linewidth]{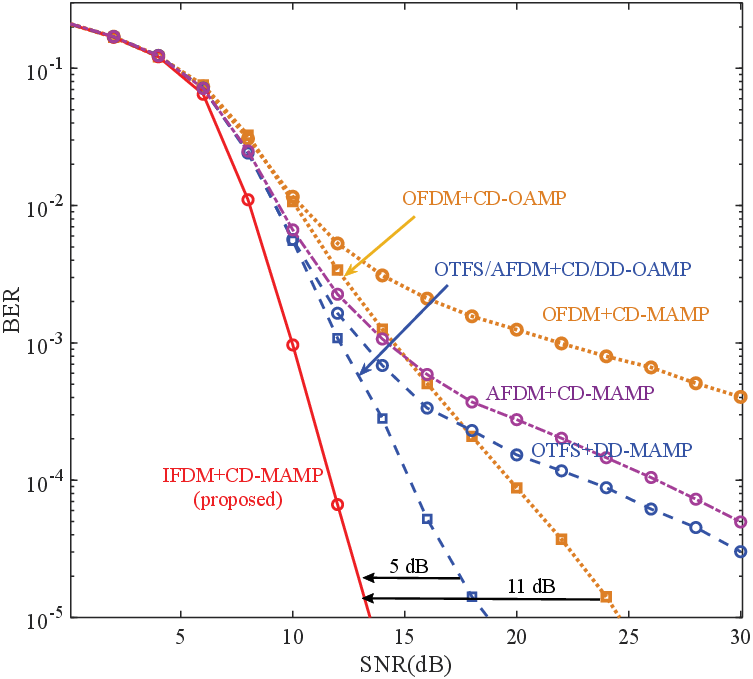}}
\subfigure[\fontsize{4.7pt}{\baselineskip}\selectfont{Running time: QPSK \!+\! MIMO (300km/h)}]{\label{Fig:timecomp}
\includegraphics[width=0.255\linewidth]{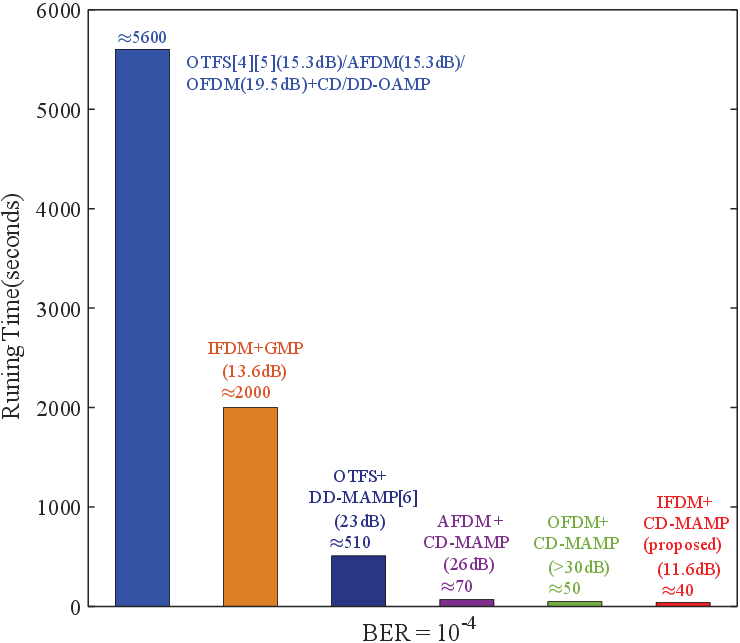}}
\vfill 
\vspace{-0.2cm}
\textbf{\subfigure[\fontsize{4.7pt}{\baselineskip}\selectfont{OFDM/OTFS/AFDM/IFDM: QPSK \!+\! MIMO (0km/h)}]{\label{Fig:BER_v0}
\includegraphics[width=0.25\linewidth]{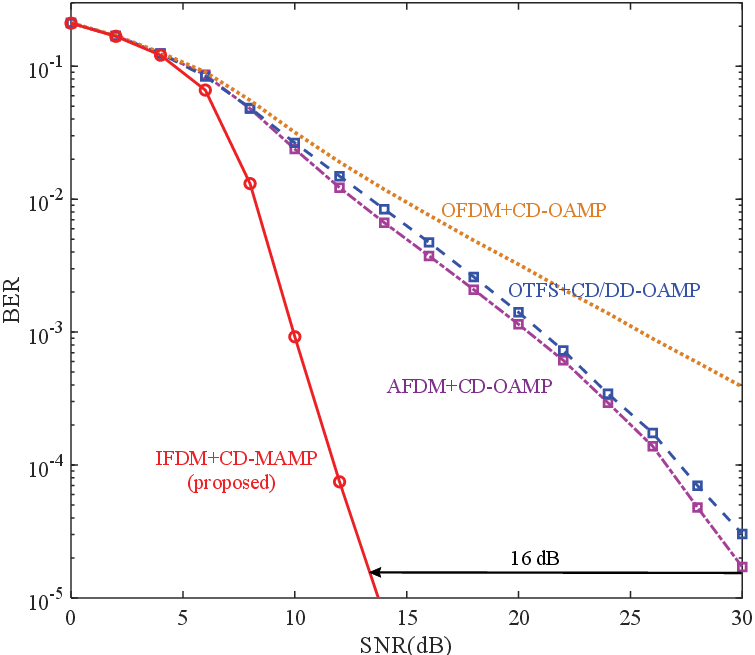}
}}
\textbf{\subfigure[\fontsize{4.7pt}{\baselineskip}\selectfont{OFDM/OTFS/AFDM/IFDM: \!QPSK+MIMO\! (500km/h)}]{\label{Fig:BER_v800}
\includegraphics[width=0.245\linewidth]{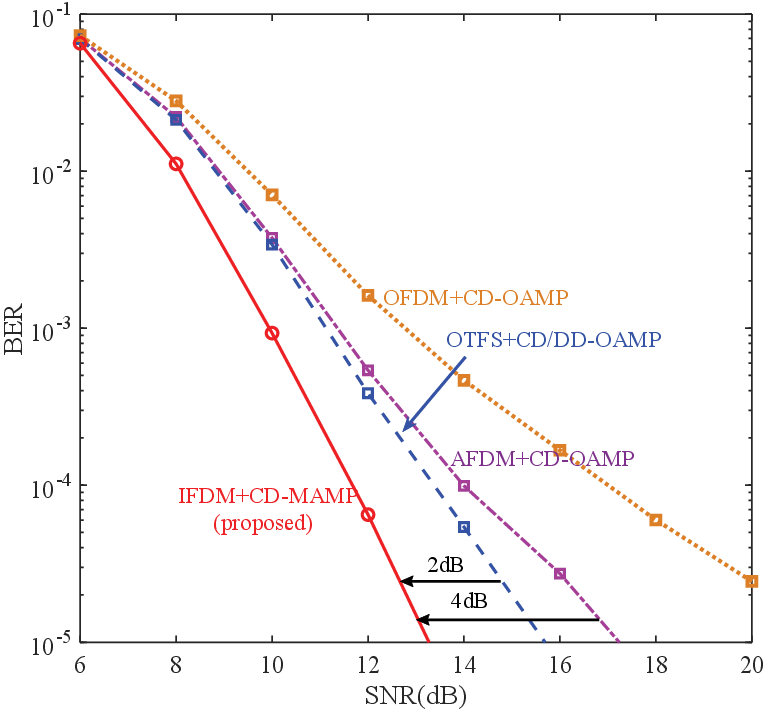}}}
\subfigure[\fontsize{4.7pt}{\baselineskip}\selectfont{\!\!\!LMMSE/GMP/OAMP/MAMP: IFDM\!+ \!SISO  + 16QAM (300km/h)}]{\label{Fig:BER_16QAM_ifdm}
\includegraphics[width=0.255\linewidth]{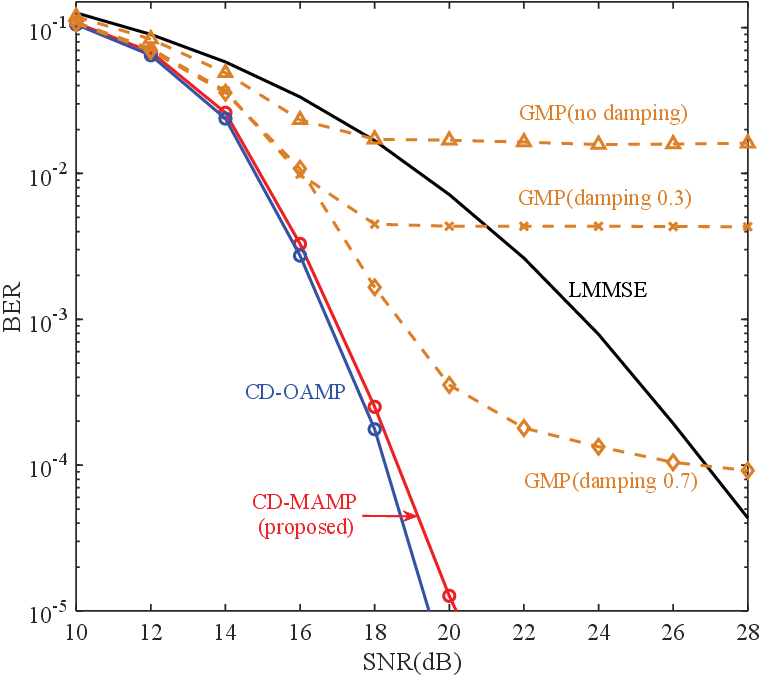}}\vspace{-0.1cm}
\caption{BER  comparisons of OFDM, OTFS, AFDM, and IFDM with different practical advanced detectors and $512$ subcarriers in SISO and $4\times4$ MIMO.}\vspace{-0.3cm}
\label{fig:subfig}
\end{figure*}

\vspace{-0.2cm}
\section{Numerical Results}\label{sec:num}
In this section, IFDM is compared to OFDM, OTFS, and AFDM with practical advanced detectors in static multipath and mobile time-varying channels. To be fair, we ensure that the multicarrier modulations use the same bandwidth, i.e., the subcarrier spacing is $\Delta f$ kHz for OTFS and OFDM and $\tfrac{\Delta f}{L}$ kHz for IFDM and AFDM. For simplicity, we consider that the carrier frequency is $4$ GHz with $\Delta f = 15$ kHz, the velocity of the device is $\{0, 300, 500\}$ km/h with a maximum Doppler frequency shift $\nu_{\text{max}}=\{0,1111,1852\}$~Hz, and the channel Doppler shift is generated by using Jakes information\cite{OTFS_GMP}. The RRC rolloff factor in the transceiver is set at $0.4$. The number of subcarrier is $N=512$ (i.e., $K=32$ and $L=16$ for OTFS), and Gray-mapped QPSK signaling and square $16$QAM signaling are employed. Furthermore, MIMO-IFDM is considered with $N_t=N_r=4$. 

To verify the benefits of IFDM, BER comparisons with OFDM, OTFS, and AFDM in SISO and MIMO using CD/DD-OAMP and CD/DD-MAMP detectors are shown in Figs.~\ref{Fig:BER_P2P_diffmod} and \ref{Fig:BER_MIMO_diffmod}. For BER curves of $10^{-5}$, IFDM with CD-MAMP can achieve about $3$ dB and $9$ dB gains with lower complexity than OTFS/AFDM\footnote{Because OTFS and AFDM have nearly identical performance with the same detectors, they are labeled together for simplicity.} with CD/DD-OAMP \cite{OTFS-OAMP} and DD-MAMP~\cite{MAMPOTFSconf}, respectively.  In MIMO, IFDM outperforms OFDM, OTFS, and AFDM by more than $5$ dB. Meanwhile, Fig. \ref{Fig:timecomp} illustrates the benefits of the extremely low complexity of IFDM with CD-MAMP.
That is, the running time of IFDM with CD-MAMP is roughly $2$ times lower than that of AFDM with CD-MAMP and more than $100$ and $10$ times lower than that of OTFS with CD/DD-OAMP and DD-MAMP, respectively.

In Figs.~\ref{Fig:BER_v0} and \ref{Fig:BER_v800}, IFDM with CD-MAMP is compared to OFDM, OTFS, and AFDM with CD/DD-OAMP in static multipath channels and very high-velocity time-varying channels ($500$ km/h). It is clear that the proposed IFDM with CD-MAMP achieves a gain of up to $16$ dB in static multipath channels and still has a gain of more than $2$ dB at $500$ km/h. This is because the larger the velocity in OTFS and AFDM, the greater the diversity gain that can be achieved.

{Furthermore, the BER comparisons of IFDM with advanced detectors for 16QAM signaling are shown in Fig. \ref{Fig:BER_16QAM_ifdm}. Note that CD-MAMP can achieve BER performances of only $0.1$ dB away from  CD-OAMP but with lower complexity. In contrast, GMP cannot use the sparse $\bf{H}$ in \eqref{Eqn:rev_mat} due to $\bf{U}$ and $\bf{U}^{-1}$, but has to be implemented using the dense $\bf{H}_{\mr{eff}}$ in \eqref{Eqn:rev_v}, resulting in up to $\mathcal{O}(N^2\mathcal{T})$ complexity. The BER of GMP is improved by incorporating damping, but it has $2\sim 5$ dB performance losses over CD-MAMP. Note that GMP requires manually adjustable damping and is thus unstable, while CD-MAMP can employ an efficient closed-form damping solution~\cite{MAMPOTFSconf}. In addition, LMMSE suffers from severe performance loss due to ignoring the \emph{a priori} information. }


\vspace{-0.35cm}
\section{Conclusion}
{
This letter investigated an IFDM modulation, which significantly outperforms OFDM, OTFS, and AFDM by considering practical advanced detectors. Particularly, the IF transform can enhance the  channels' statistical stability and enable the equivalent channel matrices to satisfy the right-unitarily invariant assumption for OAMP/MAMP detectors, ensuring that the low-complexity CD-MAMP can achieve replica MAP-optimal BER performance. Numerical results show the advantages of IFDM over OFDM, OTFS, and AFDM in SISO and MIMO in static multipath and mobile time-varying channels, making it a very competitive multicarrier modulation technique.
}


\bibliographystyle{IEEEtran}
\vspace{-0.3cm}
\bibliography{manuscript}
\end{document}